\documentclass[aps,prb,amsmath,amssymb,twocolumn,superscriptaddress,floatfix]{revtex4-2}
\usepackage{graphicx}
\usepackage{hyperref}
\usepackage{bm}
\usepackage[dvipsnames]{xcolor}
\hypersetup{pdftitle={Sr3Fe2O7},pdfauthor={Darren},pdfsubject={Frustrated magnetism},pdfdisplaydoctitle}

\def\sfo{Sr$_\text{3}$Fe$_\text{2}$O$_\text{7}$}
\def\cubic{SrFeO$_\text{3}$}
\def\two{Sr$_\text{2}$FeO$_\text{4}$}
\def\four{Sr$_\text{4}$Fe$_\text{3}$O$_\text{10}$}
\def\cue{$\mathbf{q}$}

\def\TN{\ensuremath{T_\text{N}}}

\begin{document}

\title{Rich Magnetic Phase Diagram of Putative Helimagnet \sfo}

\author{Nikita D.\ Andriushin}
\author{Justus Grumbach}

\affiliation{Institut f\"ur Festk\"orper- und Materialphysik, Technische Universit\"at Dresden, 01069 Dresden, Germany}

\author{Jung-Hwa Kim}
\affiliation{Max-Planck-Institut f\"ur Festk\"orperforschung, 70569 Stuttgart, Germany}

\author{Manfred Reehuis}
\affiliation{Helmholtz-Zentrum Berlin f\"ur Materialien und Energie, 14109 Berlin, Germany}

\author{Yuliia V.\ Tymoshenko}
\author{Yevhen A.\ Onykiienko}
\affiliation{Institut f\"ur Festk\"orper- und Materialphysik, Technische Universit\"at Dresden, 01069 Dresden, Germany}

\author{Anil Jain}
\affiliation{Max-Planck-Institut f\"ur Festk\"orperforschung, 70569 Stuttgart, Germany}
\affiliation{Solid State Physics Division, Bhabha Atomic Research Centre, Trombay, Mumbai 400085, India}
\affiliation{Homi Bhabha National Institute, Anushaktinagar, Mumbai 400094, India}

\author{W.\ Andrew MacFarlane}
\affiliation{Department of Chemistry, University of British Columbia, Vancouver, BC V6T 1Z1, Canada}
\affiliation{Stewart Blusson Quantum Matter Institute, University of British Columbia, Vancouver, BC V6T 1Z4, Canada}
\affiliation{TRIUMF, Vancouver, BC V6T 2A3, Canada}

\author{Andrey Maljuk}
\altaffiliation[Current affiliation:~]{Leibniz Institute for Solid State and Materials Research Dresden, 01069 Dresden, Germany}
\affiliation{Max-Planck-Institut f\"ur Festk\"orperforschung, 70569 Stuttgart, Germany}

\author{Sergey Granovsky}
\affiliation{Institut f\"ur Festk\"orper- und Materialphysik, Technische Universit\"at Dresden, 01069 Dresden, Germany}
%\affiliation{Faculty of Physics, M. V. Lomonossow Moscow State University, Moscow 119991, Russia}

\author{Andreas Hoser}
\affiliation{Helmholtz-Zentrum Berlin f\"ur Materialien und Energie, 14109 Berlin, Germany}

\author{Vladimir Pomjakushin}
\affiliation{\mbox{Laboratory for Neutron Scattering and Imaging, Paul Scherrer Institut, 5232 Villigen, Switzerland}}

\author{Jacques Ollivier}
\affiliation{Institut Laue-Langevin, 71 Avenue des Martyrs, CS 20156, 38042 Grenoble CEDEX 9, France}

\author{Mathias Doerr}
\affiliation{Institut f\"ur Festk\"orper- und Materialphysik, Technische Universit\"at Dresden, 01069 Dresden, Germany}

\author{Bernhard Keimer}
\affiliation{Max-Planck-Institut f\"ur Festk\"orperforschung, 70569 Stuttgart, Germany}

\author{Dmytro S.\ Inosov}
\email{dmytro.inosov@tu-dresden.de}
\affiliation{Institut f\"ur Festk\"orper- und Materialphysik, Technische Universit\"at Dresden, 01069 Dresden, Germany}
\affiliation{W\"urzburg-Dresden Cluster of Excellence on Complexity and Topology in Quantum Matter\,---\,ct.qmat, Technische Universit\"at Dresden, 01069 Dresden, Germany}

\author{Darren C.\ Peets}
\email{darren.peets@tu-dresden.de}
\affiliation{Institut f\"ur Festk\"orper- und Materialphysik, Technische Universit\"at Dresden, 01069 Dresden, Germany}
\affiliation{Max-Planck-Institut f\"ur Festk\"orperforschung, 70569 Stuttgart, Germany}

\begin{abstract}

\noindent
The cubic perovskite \cubic\ was recently reported to host hedgehog- and skyrmion-lattice phases in a highly symmetric crystal structure which does not support the Dzyaloshinskii-Moriya interactions commonly invoked to explain such magnetic order.  Hints of a complex magnetic phase diagram have also recently been found in powder samples of the single-layer Ruddlesden-Popper analogue \two, so a reinvestigation of the bilayer material \sfo, believed to be a simple helimagnet, is called for. Our magnetization and dilatometry studies reveal a rich magnetic phase diagram with at least six distinct magnetically ordered phases and strong similarities to that of \cubic. In particular, at least one phase is apparently multiple-\cue, and the \cue s are not observed to vary among the phases. %Far from being a simple helimagnetic baseline for comparison to \cubic, \sfo\ is likely to host an array of highly nontrivial magnetic phases of its own. 
Since \sfo\ has only two possible orientations for its propagation vector, some of the phases are likely exotic multiple-\cue\ order, and it is possible to fully detwin all phases and more readily access their exotic physics.

\end{abstract}

%%%%%%%%%%%%%%%%%%%%%%%%%%%%%%%%%%%%%%%%%%%%%%%%%%%%%%%%%%%%%%
\maketitle
%%%%%%%%%%%%%%%%%%%%%%%%%%%%%%%%%%%%%%%%%%%%%%%%%%%%%%%%%%%%%%

\section{Introduction}
Helimagnets are a special class of materials that realize noncollinear long-range-ordered magnetic structures in the form of proper-screw spin helices~\cite{Tokura2010,Braun2012,Kimura2012,Nagaosa2013}, and fall into two general categories based on their underlying crystal structures and correspondingly the mechanism of stabilization of the magnetic structure: (1)~helimagnets in compounds having noncentrosymmetric crystal structures or in which Dzyaloshinskii-Moriya (DM) interactions are possible (Dzyaloshinskii-type helimagnets)~\cite{Dzyaloshinsky1958, Moriya1960, Dzyaloshinskii1964, Dzyaloshinskii1965, Roessler2006, Togawa2016}; and (2)~helimagnets in compounds having centrosymmetric crystal structures with inversion symmetry at the midpoint between consecutive magnetic atoms (Yoshimori-type helimagnets)~\cite{Yoshimori1959, Kaplan1959, Villain1959}. In the former, broken inversion symmetry in the underlying crystal lattice allows relativistic DM interactions which twist the magnetic moments, and the propagation vector and the pitch angle are controlled by the Dzyaloshinskii vector $\mathbf{D}$. In the latter, the chirality must emerge due to a spontaneous chiral symmetry breaking.  Helical order is described by an ordering vector \cue, which is usually incommensurate with the lattice.

%Under a magnetic field perpendicular to the helical axis, the helical magnetic state in noncentrosymmetric helimagnets continuously evolves to a field-induced ferromagnetic state via a soliton lattice state.  On the other hand, symmetric helimagnets are expected to first undergo a discontinuous transition into a fan structure, followed by a continuous approach to the field-induced ferromagnetic state. The collective spin excitations in a noncentrosymmetric helimagnet are gapped due to anisotropy, while a symmetric helimagnet possesses three Goldstone modes (\Cue=0,~$\pm q$).

Very rarely, materials with helimagnetic order can exhibit more-complex topologically nontrivial multiple-\cue\ structures such as skyrmion or hedgehog lattices~\cite{Muehlbauer2009}, in which multiple helical orders propagating along different directions couple to produce a more complex noncoplanar ordering pattern. This leads to a lattice of topological defects, resembling spin vortices or hedgehogs, in which the spin direction winds around a central line or point~\cite{Finocchio2016,Fert2017, EverschorSitte2018}.  The topological protection makes these entities relatively robust, and in film form, skyrmions have been investigated for use in magnetic memory.  However, despite these spin structures being relatively robust, the ordered phase in which they occur is typically only found in a narrow bubble of temperature and field~\cite{Ishikawa1984}. Several materials are known to host two-dimensional lattices of skyrmions, primarily the structurally-chiral Cu$_\text{2}$OSeO$_\text{3}$~\cite{Seki2012c, Seki2012b}, MnSi~\cite{Muehlbauer2009}, and closely-related binary materials forming in the same B20 structure as MnSi~\cite{Yu2010,Muenzer2010,Yu2011}, in which the skyrmion phases arise as a consequence of DM interactions and the chiral symmetry of the lattice. Within the past four years, a handful of {\slshape centro\-symmetric} materials were reported to host skyrmion-lattice phases, offering a completely different route to skyrmion-lattice physics based on distinct underlying interactions.  The first such materials were \cubic~\cite{Ishiwata2020} and Gd$_2$PdSi$_3$~\cite{Kurumaji2019,Zhang2020}, and the former is also proposed to host a three-dimensional hedgehog-lattice phase.  In contrast to the small multiple-\cue\ bubbles typically seen in noncentrosymmetric materials, multiple-\cue\ phases in centrosymmetric materials may occupy much of the magnetic phase diagram.  The centrosymmetric skyrmion materials may also host exotic multiple-\cue\ orders beyond skyrmion- and hedgehog-lattice phases, for instance the vortex state with stripes of topological charge recently reported in GdRu$_2$Si$_2$~\cite{Wood2023}.  

The cubic perovskite \cubic\ has a particularly intriguing $H$--$T$ phase diagram with at least five distinct magnetic phases for $\mathbf{H}\parallel[111]$ alone~\cite{Ishiwata2011}.  Two of these five phases have been identified~\cite{Ishiwata2020}: A double-\cue\ skyrmion-lattice phase and a quadruple-\cue\ phase producing a three-dimensional lattice of hedgehogs and anti-hedgehogs. \cubic\ is the three-dimensional ($n=\infty$) member of a Ruddlesden-Popper family of layered materials, including the single-layer analogue \two, bilayer compound \sfo, and triple-layer material \four~\cite{Fossdal2004}, of which single crystals of only \sfo\ have been grown.  This latter material has been reported to be a helimagnet with a slightly elliptical helix~\cite{Peets2013, Kim2014} whose spins lie perpendicular to the tetragonal [110] direction; its $(\xi\,\xi\,1)$ propagation vector (with $\xi=0.142$ and antiferromagnetic stacking of bilayers) is the quasi-two-dimensional analogue of the $(\xi\,\xi\,\xi)$ in cubic \cubic\ with $\xi=0.128$~\cite{Reehuis2012}. This close similarity is particularly remarkable given that \sfo\ is an insulator below 330\,K while \cubic\ is a metal. The insulating behavior arises from freezing of a checkerboard charge modulation which breaks the symmetry between adjacent Fe ions~\cite{Kim2021}. The associated lowering of the lattice symmetry could in principle allow DM interactions, but the small changes in atomic positions and highly similar propagation vectors suggest that DM interactions play no significant role. \two\ was very recently reported to exhibit elliptical {\itshape cycloidal} order with the similar \cue\ vector $(\xi\,\xi\,0)$, $\xi=0.137$~\cite{Adler2022}, while the magnetism in \four\ has not been reported. The work on \two\ identified a transition within the ordered phase at 10\,K, a shoulder in the magnetization at 30\,K under a 3.5\,T field, a spin-flop transition near 5\,T, and a transition to ferromagnetic order between 5 and 8\,GPa, indicating a complex magnetic phase diagram. The complexity found in \cubic\ and \two\ suggests that the $H$--$T$ phase diagram of \sfo\ should be investigated in detail.

In this work, we explore the magnetic phase diagram of \sfo\ using magnetization and dilatometry measurements, finding a similarly rich phase diagram. The parallels with \cubic\ suggest exotic multiple-\cue\ order, and we are able to constrain the possibilities for several phases.

\section{Experimental}

Large single crystals of \sfo\ were prepared by floating-zone growth as described previously~\cite{Maljuk2004,Peets2013}.  The oxygen content was maximized by annealing under 5\,kbar of oxygen while gradually cooling from 450\,$^\circ$C~\cite{Kim2021}, or for some powder samples by annealing at 6\,kbar at 550\,$^\circ$C, and was verified to be O$_{>6.99}$ by thermogravimetric analysis and structure refinements.  High sample quality was confirmed by diffraction --- previous synchrotron powder diffraction found these samples to be phase pure\,\cite{Kim2021}, resonant x-ray diffraction found rocking curves typically 0.05--0.10$^\circ$ wide on smaller crystals, while neutron rocking curves on larger crystals were 1--2$^\circ$ wide.
The sample for neutron powder diffraction was prepared by standard solid-state synthesis, and contained \cubic\ as an impurity phase.

Magnetization measurements were performed by vibrating sample magnetometry (VSM) in a Quantum Design Magnetic Property Measurement System (MPMS-VSM) or in a Cryogenic Ltd.\ Cryogen-Free Measurement System (CFMS) using the VSM module, in zero-field-warming, field-cooled-cooling, and field-cooled-warming conditions.  The ac susceptometry option was used for frequency-dependent measurements in a 0.5\,Oe ac field.  Four- or five-quadrant $M$-$H$ loops were measured at several temperatures in the CFMS. The single crystals were mounted to either a plastic (CFMS) or quartz rod (MPMS) sample holder using GE varnish.  

Specific heat measurements were performed in a Quantum Design Physical Property Measurement System (PPMS) with the sample secured using Apiezon N grease.

Dilatometry measurements were performed using a tilted-plate capacitive dilatometer with a sensitivity to relative length changes of $\sim 10^{-7}$~\cite{Rotter1998}, which was mounted on an Oxford Instruments $^4$He flow cryostat equipped with a superconducting magnet capable of fields up to 10\,T.  The sweep rate of the magnetic field was chosen to be between 0.05\,T/min and 0.25\,T/min. For accurate monitoring and control of the dilatometer and sample temperature we used a Cernox thermometer attached to the dilatometer cell close to the sample.  Measurements of magnetostriction and thermal expansion were made on single crystals for length changes parallel or perpendicular to the crystallographic [110] or [001] directions for magnetic fields oriented in both of these directions. The longitudinal and transverse components of the striction tensor found in this way allow the distortions and volume effects of the crystal lattice to be calculated. This allows one to identify all magnetic transitions accompanied by lattice effects through dilatometry, which can hint at modifications of the magnetic structure.

\begin{figure*}[htb]
  \includegraphics[width=\textwidth]{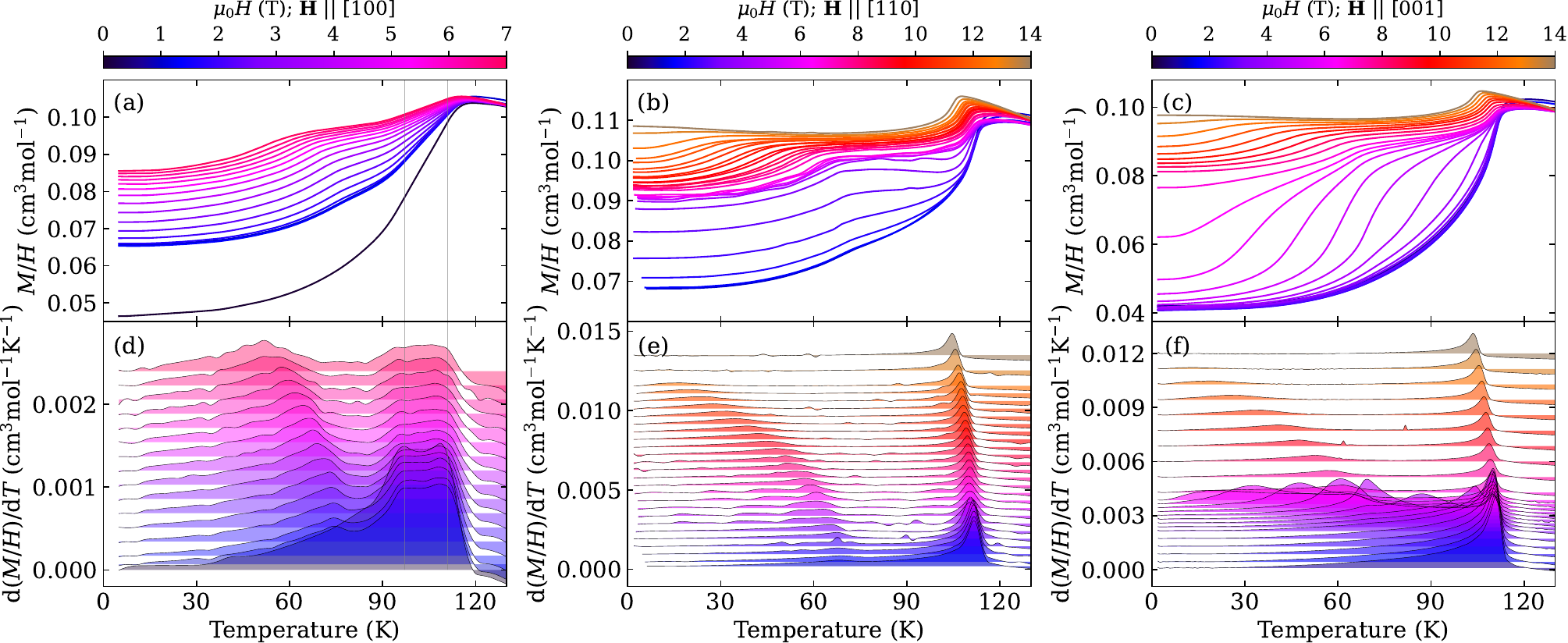}
  \caption{\label{MT1}Temperature dependent field-cooled magnetic susceptibility $\mathbf{M/H}$. Plotted for applied magnetic fields along (a) [100], (b) [110], and (c) [001] directions. The respective derivatives are plotted in panels (d-f), in which the datasets have been offset vertically for clarity. In (a) and (d), gray vertical lines show where the peaks in the derivative occur at low field.}
\end{figure*}

Single-crystal neutron diffraction was performed on the E5 diffractometer at the BER-II reactor at the Helmholtz-Zentrum Berlin (HZB), Germany. The wavelength 2.38\,\AA\ was selected using the (002) reflection from a pyrolytic graphite (PG) monochromator, and higher-order contamination ($\lambda/2$) was prevented through the use of a PG filter. A position-sensitive $^3$He detector of dimension 90$\times$90\,mm$^2$ was used. Samples were mounted in four-circle geometry on a closed-cycle refrigerator, and collimators and slits were set such that each sample was fully illuminated. Data were integrated using the {\sc racer} program~\cite{Wilkinson1988}, which uses the parameters describing the shape of strong peaks to improve the precision in the description of weaker ones, minimizing the relative standard deviation.  Further measurements in fields applied along [110] and [001] were performed at beamline E4 at the BER-II reactor at the HZB, using a 2D detector and neutrons of wavelength 2.437\,\AA.  Powder neutron diffraction was measured with 1.8857-\AA\ neutrons in applied magnetic fields up to 6\,T at the HRPT beamline at the Paul Scherrer Institute (PSI), Villigen, Switzerland, and up to 6.5\,T at the E6 diffractometer at the BER-II reactor at the HZB using 2.42-\AA\ neutrons selected by a PG monochromator.  

The effectiveness of detwinning the magnetic order (i.e.\ selecting a single-domain magnetic state) in a field $\mathbf{H}\parallel[110]$ was checked using the IN5 time-of-flight beamline at the Institute Laue-Langevin (ILL), Grenoble, France, using a neutron wavelength of 4.8\,\AA. The sample was cooled to 1.8\,K in the maximum 2.5-T field possible at this beamline, then measured in zero field at this temperature.  Data were integrated from $-0.05$ to 0.05\,meV to capture elastic scattering, while out-of-plane momentum integration was set to $\pm0.04$ reciprocal lattice units (r.l.u.).  

Throughout this paper, crystal orientations refer to the high-temperature tetragonal $I4/mmm$ cell, rather than the doubled charge-ordered $Bmmb$ cell. The helical propagation vector is \mbox{$(\xi~\pm\xi~1)$} in the tetragonal cell, or $(\sqrt{2}\xi~0~1)$/$(0~\sqrt{2}\xi~1)$ in the $Bmmb$ cell. The charge order has a correlation length along [001] on the order of a unit cell~\cite{Kim2021}, but the magnetic order produces sharp peaks in neutron powder diffraction~\cite{Kim2014}, so a magnetic domain must include a considerable number of structural domains and feel an effectively tetragonal lattice.

To further characterize the magnetic order, muon spin rotation/relaxation measurements ($\mu$SR) using positive muons ($\mu^+$) were performed on a single crystal mounted on the low-background insert of a helium-flow cryostat in the LAMPF spectrometer installed at the M15 beamline at TRIUMF, in Vancouver, Canada. In this setup, muons that do not stop in the sample are vetoed with a high efficiency. The crystalline $c$ axis was parallel to both the incident muon beam and its spin. Decay positrons were detected in a pair of scintillation detectors up- (B) and downstream (F) of the sample. The muon spin polarization is then monitored by the experimental asymmetry of the count rates, $A = (B-F)/(B+F)$. For more details, see our earlier report\,\cite{MacFarlane2014}. 

\begin{figure}[t]
  \includegraphics[width=\columnwidth]{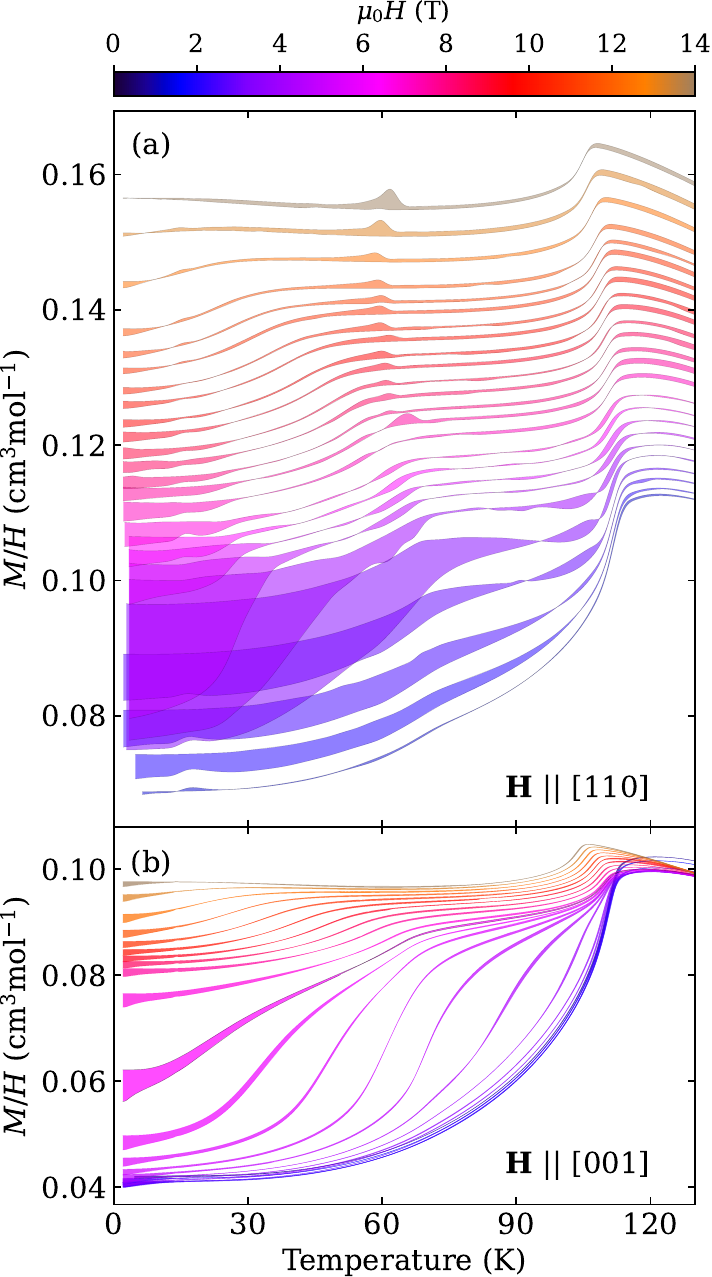}\vspace{-5pt}
  \caption{Comparison of field-cooled and zero-field-cooled magnetization data. Plotted for selected fields parallel to (a)~[110] and (b)~[001]; shading indicates the difference. The datasets in (a) have been offset vertically for clarity.\label{ZFC}}
\end{figure}

\section{Magnetization}

Magnetization measurements were performed on \sfo\ as a function of field $\mathbf{H}$ and temperature $T$ for applied magnetic fields along [100], [110], and [001]; field-cooled data for all three directions are shown in Fig.~\ref{MT1} together with their derivatives.  The first transition encountered on cooling, which we refer to as \TN, is at roughly 111\,K, consistent with previous reports. However, it is immediately clear that there is an additional transition within the magnetically ordered phase in field for all field orientations, starting around 70\,K at low field and moving to lower temperature as field is increased.  There is also some evidence, most clearly seen in the derivatives, that the first transition encountered may be split. It is also striking that the magnetization at low temperatures changes drastically in field.

Zero-field-cooled magnetization data are presented in Fig.~\ref{ZFC}(a) for $\mathbf{H}\parallel[110]$ and in Fig.~\ref{ZFC}(b) for \mbox{$\mathbf{H}\parallel[001]$}. ZFC data were not collected for $\mathbf{H}\parallel[100]$. The features seen in the FC data are also visible here, but several new features appear. 

\begin{figure}[t]
  \includegraphics[width=\columnwidth]{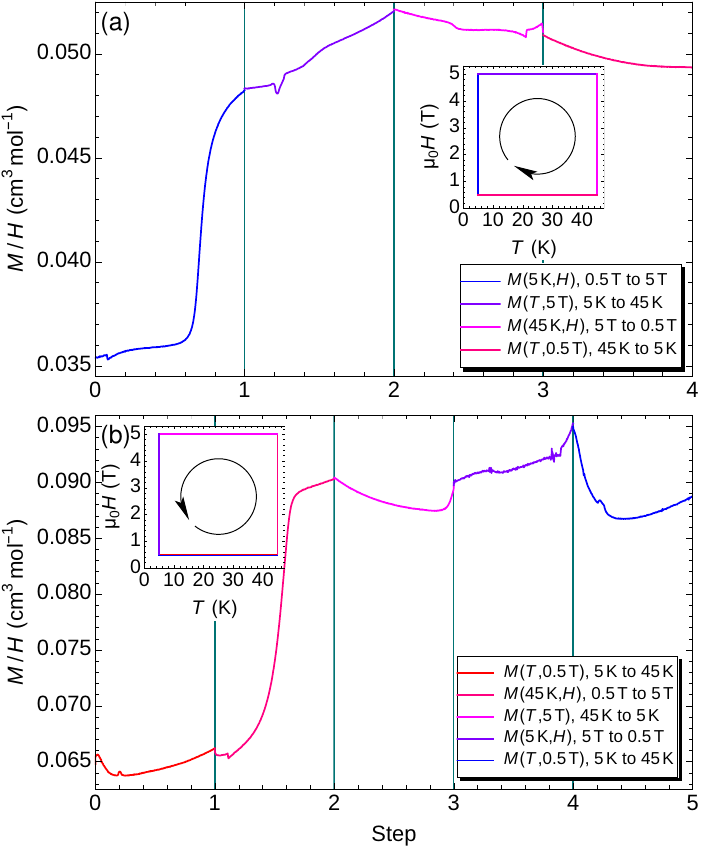}
  \caption{Demonstration of field training. Circuits through the $H$--$T$ phase diagram at low temperature and low field, showing the effect of field training. Insets show the paths taken through the phase diagram.\label{loop}}
\end{figure}

\begin{figure*}
  \includegraphics[width=\textwidth]{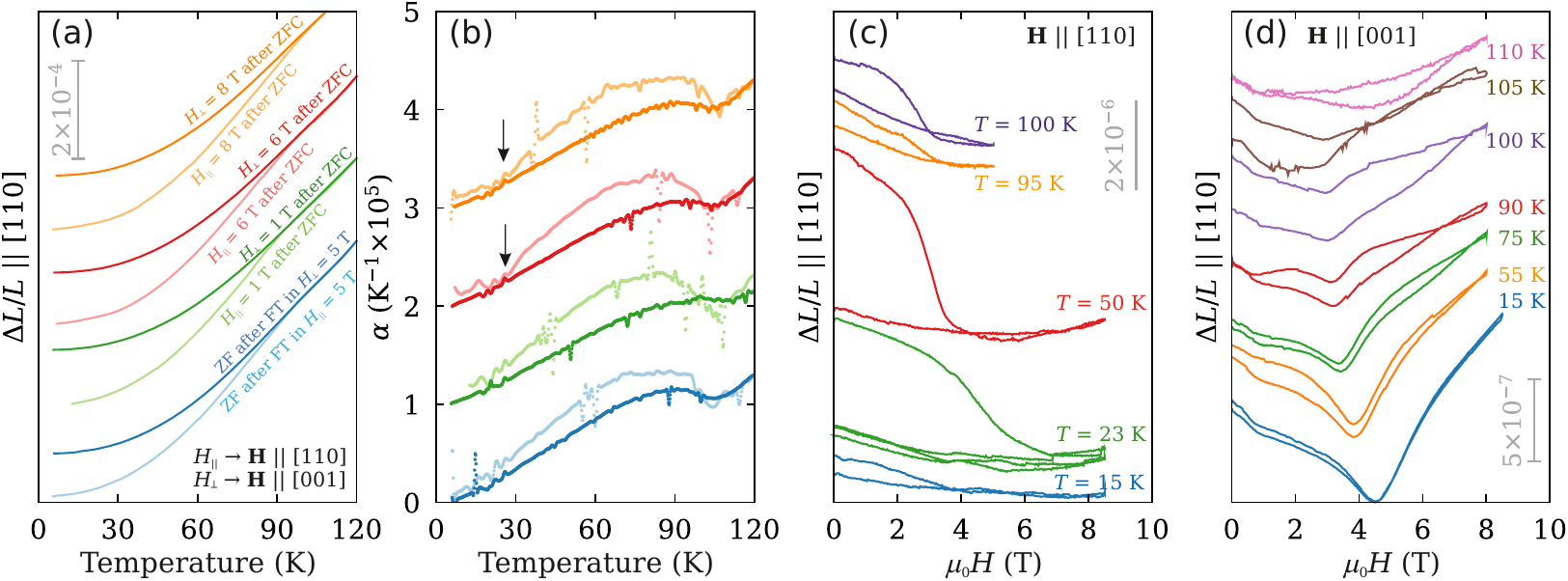}
  \caption{\label{DLT}Dilatometry results on \sfo. (a) Zero-field-cooled and zero-field (ZF) field-trained [110] thermal expansion data taken on warming for transverse ($\mathbf{H}\parallel[001]$, bold lines) and longitudinal ($\mathbf{H}\parallel[110]$, pale lines) field conditions at selected fields. (b) Linear expansion coefficient $\alpha$, \textit{i.e.} derivative of the data in panel (a). Offset is applied for visual clarity. (c), (d) Longitudinal and transverse magnetostriction data measured after ZFC.}
\end{figure*}

The ZFC data diverge significantly from the FC data below $\sim$30\,K for intermediate [110] fields, as shown in Fig.~\ref{ZFC}(a), indicating a freezing of spin components or domains that would otherwise be field-trained by a sufficiently strong field.  Lower fields are not strong enough to field-train the magnetic order, and higher fields suppress the ZFC-FC splitting to lower temperature and reduce the splitting. In Fig.~\ref{loop}, two circuits are shown which take opposite paths through the $H$--$T$ phase diagram, starting from 5\,K and 0.5\,T under zero-field-cooled conditions, and neither returns to its initial magnetization value.  In both circuits a single-domain state is obtained.  In both cases, a large step is seen on increasing the field to 5\,T, but our $M(H)$ data (shown below in Fig.~\ref{MH}) indicate that if we stayed at 5\,K, a decreasing field would follow the same curve, since 14\,T is insufficient to detwin the magnetic order for temperatures up to at least 10\,K.  The circuits in Fig.~\ref{loop} exceed this temperature at high field, detwinning the magnetic order.

A smaller difference between ZFC and FC data is also observed for fields $\mathbf{H}\parallel[001]$. We do not see evidence of field training into a single-domain state for this orientation, so it is not clear what is being frozen or trained.  

Above $\sim$6\,T, a peak appears around 60\,K in the ZFC data for $\mathbf{H}\parallel[110]$, which disperses to slightly higher temperatures as the field is increased. This enhanced response to the applied field suggests a phase transition, likely out of a frozen-in low-field state. 

Differences between ZFC and FC data also appear at some phase transitions, where they most likely arise from hysteresis between cooling (FC) and warming (ZFC) data.  Similar hysteresis has been seen previously in the 60 and 110\,K transitions in \cubic~\cite{Ishiwata2011}.  An additional dip visible around 14\,K in all data taken on warming is associated with a change in the cryostat's cooling mode and does not arise from the sample.  

Taken together with the FC data, these ZFC data make it clear that \sfo\ has a rather complex $H$--$T$ phase diagram.  

\section{Dilatometry}

Dilatometry experiments assess thermal expansion  --- the change in the unit cell as a function of temperature --- as well as expansions due to other parameters such as magnetic field. Magnetoelastic coupling induces forced magnetostriction upon the application of an external magnetic field, while below the ordering temperature, spontaneous magnetostriction can manifest. We applied magnetic fields in two configurations: parallel to the length measurement direction (longitudinal field) and orthogonal to it (transversal field). In contrast to forced magnetostriction in the paramagnetic phase, magnetostriction below the N{\'e}el temperature exhibits anisotropy in helimagnets, usually leading to a divergence of the transversal and longitudinal datasets in the ordered phase. This effect is distinct from magnetic detwinning and stems from the inherent anisotropy of magnetoelastic coupling.

Changes of the sample length along the [110] direction caused by strong magnetoelastic coupling were studied in fields along the [110] direction (longitudinal) and the [001] direction (transversal). Thermal expansion data were recorded upon increasing temperature after zero-field cooling (1, 6, and 8\,T) and after field training (0\,T) and are shown in Fig.~\ref{DLT}(a). Their derivatives, representing the coefficient of linear expansion $\alpha$, are presented in Fig.~\ref{DLT}(b). The measurement curves are nearly parallel for both directions but with a pronounced difference in absolute value, and they only converge at the magnetic ordering temperature $\TN=110$\,K. From the absolute values in Fig.~\ref{DLT}(a), lattice parameter changes on the order of $5\times10^{-5}$ can be estimated. Assuming a constant unit-cell volume, the sample expansion in the [110] direction upon entering the ordered phase would correspond to a contraction of the $c$ lattice parameter. 

In the coefficient of linear expansion $\alpha$, phase transitions are indicated by kinks; the ordering temperature also manifests clearly in this way. At lower temperatures, a first kink in $\alpha(T)$ for 6 and 8\,T at 25\,K correlates with the recovery seen in the zero-field-cooled magnetization for fields above 4\,T in Fig.~\ref{ZFC}. This transition is identified with arrows in Fig.~\ref{DLT}(b). In addition, others are visible for the longitudinal $\mathbf{H}\parallel[110]$ case for all measured external fields. Anomalies at 63--70\,K correspond to the transitions seen in this temperature range in the magnetization.

\begin{figure*}
  \includegraphics[width=\textwidth]{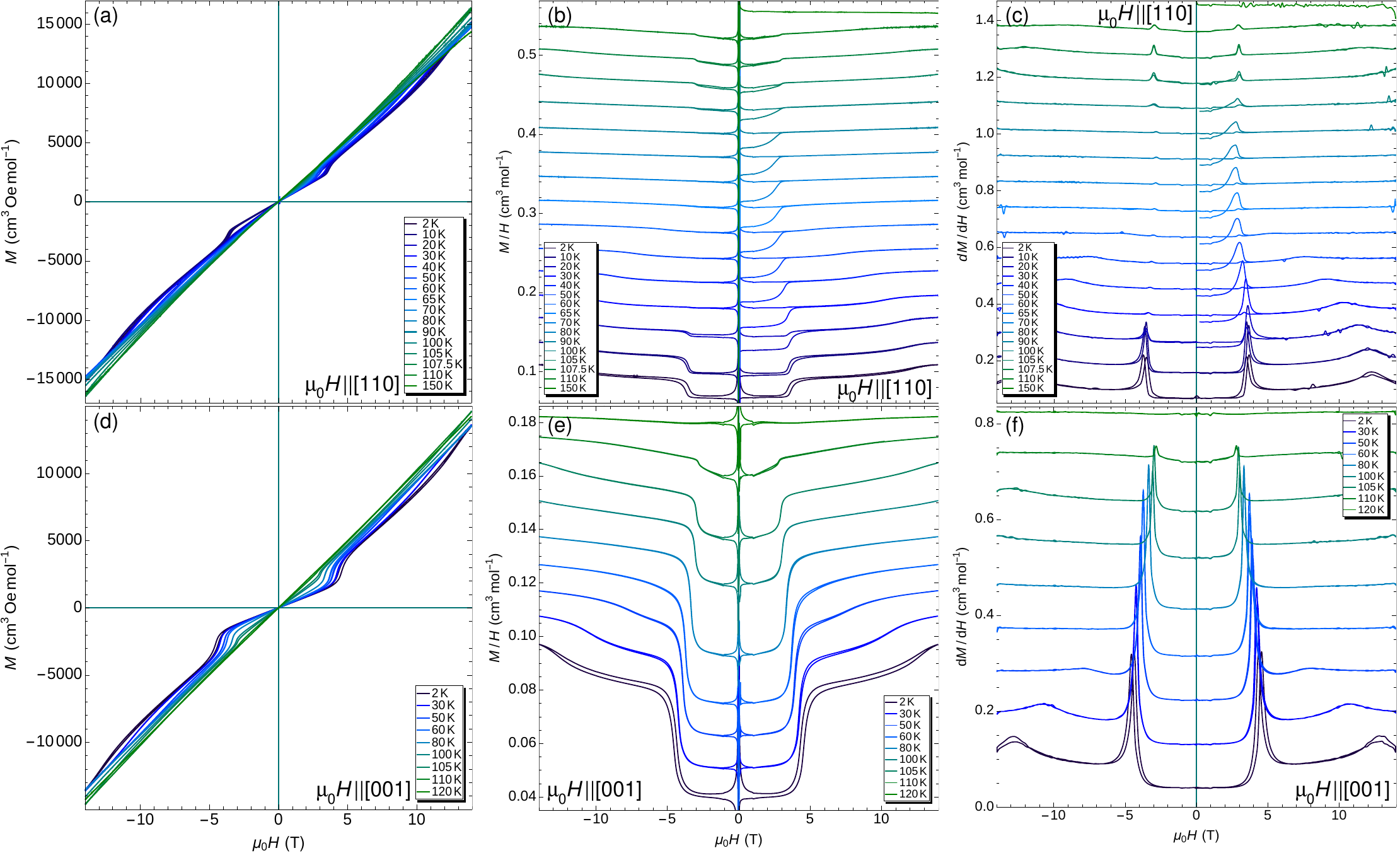}
  \caption{\label{MH}Field-dependent magnetization data at selected temperatures. Plotted as (a) $M(H)$ and (b) $M(H)/H$ for $\mathbf{H}\parallel[110]$ and as (d) $M(H)$ and (e) $M(H)/H$ for $\mathbf{H}\parallel[001]$. Derivatives are plotted for (c) $\mathbf{H}\parallel[110]$ and (f) $\mathbf{H}\parallel[001]$.  Vertical offsets have been added to the data in panels (b), (c), (e), and (f) for clarity.}
\end{figure*}

Field-dependent magnetostriction measurements in the longitudinal setup ($\mathbf{H}\parallel[110]$) are shown in Fig.~\ref{DLT}(c). The first field sweep in each case is associated with a clear irreversible length reduction of about $2\times10^{-6}$ at 2.9--4.5\,T , while in all following field sweeps at the same temperature the increasing-field curves nearly exactly track the decreasing-field curves with no irreversibility. This can be explained with a possible field training of the magnetic structure, for instance a domain-selection process. A strong kink is observed around 4\,T, consistent with the transition in the magnetization results and the detwinning field observed here.  The discussed anomalies and phase transitions agree quite well with the magnetization data.

Measurements of the magnetostriction in the transverse setup ($\mathbf{H}\parallel[001]$) in Fig.~\ref{DLT}(d) show a clear transition around 0.7\,T which does not have obvious signatures in the magnetization. These transitions can thus be concluded to be lattice-driven or -influenced. 

\section{Field Training\label{sec:detwin}}

\begin{figure}[t]
  \includegraphics[width=\columnwidth]{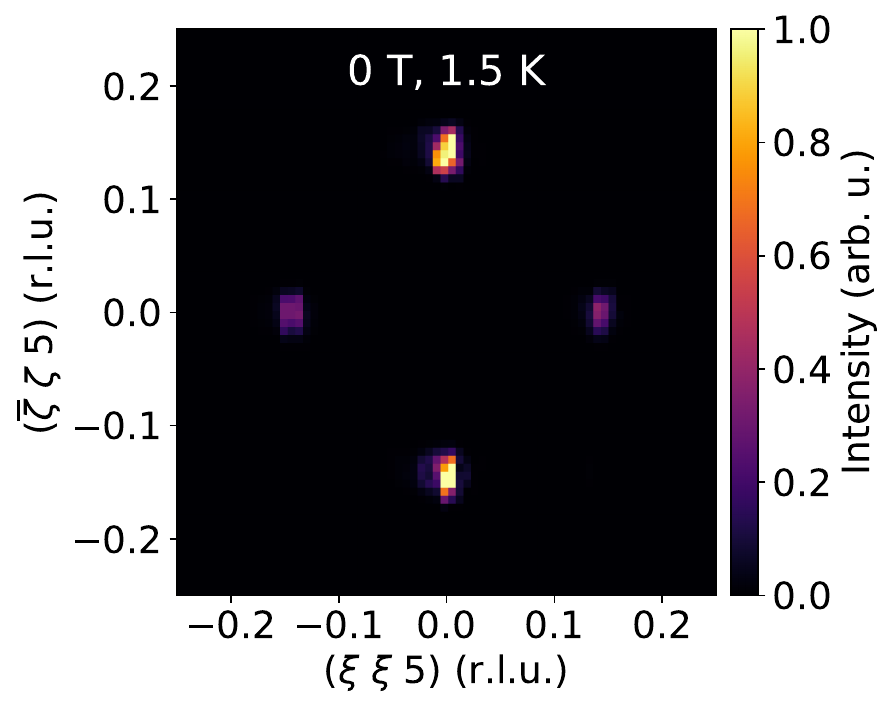}
  \caption{\label{NS}Effect of domain selection on magnetic Bragg peaks. Elastic neutron scattering intensity in the magnetic satellites around the structurally forbidden (005) reflection in zero field at 1.5\,K after cooling in a field $\mu_0\mathbf{H}\parallel[1\overline{1}0]$ of 2.5\,T.}
\end{figure}

In \sfo\ there are two equivalent directions for the helical propagation vector, $(\xi~\pm\xi~1)$, and the sample is expected to form a multidomain state with roughly equal contributions of both if cooled in zero field.  As was discussed for instance in connection with ZnCr$_2$Se$_4$~\cite{InosovZnCrSe2020}, it is often possible to detwin helical magnetism by applying a magnetic field perpendicular to the plane of the spins corresponding to one of the equivalent propagation vectors. The helix associated with that propagation vector can readily add a third component along the field to become conical with minimal impact on its ordered components, but other orientations of the helix are destabilized.  The single-domain state thus prepared usually remains stable when the field is removed, due to the energy cost of nucleating domain walls.  In \sfo\ detwinning requires a field along [110].  To test for detwinning behavior and determine the required field strength, we measured magnetization as a function of field (4-quadrant $M$-$H$ loops) in this field orientation.  Selected data are plotted in Fig.~\ref{MH}(a), and more temperatures are plotted as $M/H$ in Fig.~\ref{MH}(b).  Derivatives are plotted in Fig.~\ref{MH}(c).  As can be seen, there is a clear transition around 3\,T, for both positive and negative field sweep directions, closely resembling the spin-flop transition reported recently in powder samples of the single-layer analogue \two~\cite{Adler2022}.  At most temperatures this transition is accompanied by an irreversible detwinning transition\,---\,the $M/H$ values found before first reaching this field cannot be obtained again by field sweeps alone. This magnetic detwinning was verified by neutron scattering, as shown in Fig.~\ref{NS}. This sample was cooled in a field of 2.5\,T applied along the [1$\overline{1}$0] direction, then measured in zero field. The magnetic reflections along the field were $\sim$3 times more intense than those perpendicular to the field, consistent with the partial detwinning expected for a field somewhat below the 3--4\,T transition. The ability to detwin the magnetic order means that besides field-cooled and zero-field-cooled conditions, it is possible to measure the sample in its \emph{single-domain state} obtained by field training.  

\begin{figure}[tb]
  \includegraphics[width=\columnwidth]{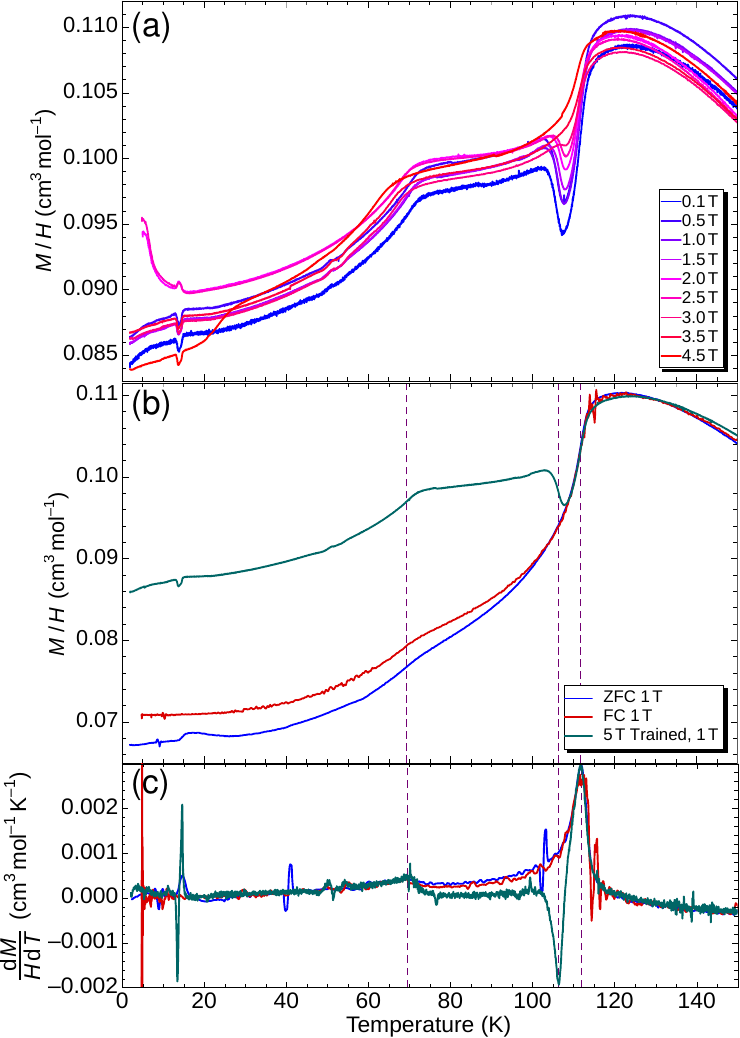}
  \caption{\label{training}Effect of field training on magnetization. Magnetization data collected under field-trained conditions with $\mathbf{H}\parallel[110]$, to prepare a single-domain state. (a)~Field-trained data measured on warming at several applied fields. At 2 and 2.5\,T, the sample was cooled in 5\,T to base temperature; for the other datasets, the 5\,T field was reduced to 0\,T at 50\,K before continuing to cool to base temperature. (b)~A comparison of FC, ZFC, and field-trained data measured in $\mu_0H=1$\,T. (c) Derivatives of curves in (b).}
\end{figure}

\begin{figure}[b]
  \includegraphics[width=\columnwidth]{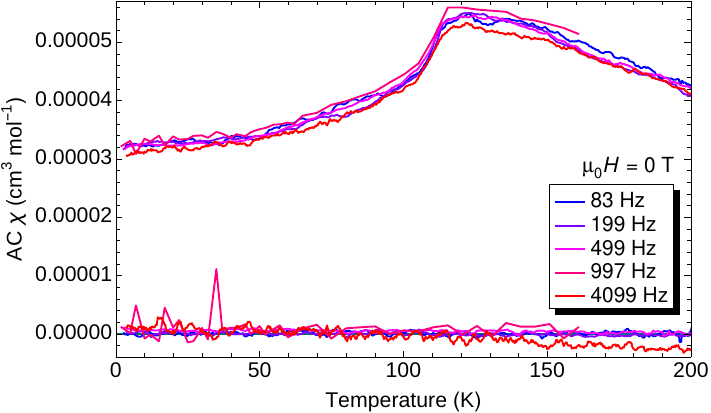}
  \caption{\label{ACchi}ac susceptibility of \sfo. Real (upper) and imaginary (lower) components of the temperature-dependent ac susceptibility of \sfo\ at zero applied field, for several frequencies. There is no clear feature in the loss, and no evidence of frequency dependence.}
\end{figure}

Knowing that 3--4\,T is sufficient to detwin the magnetism at most temperatures, we took additional data with a third field history. For these field-trained data, shown in Fig.~\ref{training}(a), the sample was cooled from well above \TN\ in a field of 5\,T, typically to a temperature of $\sim$50\,K, before cooling to base temperature in zero field, upon which the sample was measured on warming in an applied field. A comparison of ZFC, FC, and field-trained data at 1\,T is shown in Fig.~\ref{training}(b), and the derivatives in Fig.~\ref{training}(c). The field-trained data are vastly different from the other datasets over most of the temperature range, indicating detwinning of the magnetism. The field-trained curves rejoin the other field histories in a sharp transition roughly 7\,K below \TN.  In tests of the detwinning, we found that detwinning was preserved if we warmed to temperatures below this transition and cooled again, but detwinning was lost if we warmed into this transition.  Such a transition would be explainable as either relaxation through fluctuations, or through the system entering a small bubble of a multiple-\cue\ phase just below \TN.  ac susceptometry curves (Fig.~\ref{ACchi}) closely follow the dc magnetization curves, do not shift with frequency, and do not have clear features in the imaginary component, excluding fluctuations, and the field dependence in the $M$--$H$ loops does not suggest improved detwinning at higher fields, so this is most likely a multiple-\cue\ phase.  Since there are only four possible \cue\ orientations in this system\,---\,$\pm$(110) and $\pm$($\overline 1$10)\,---\,assuming the \cue\ itself does not change and no component parallel to \cue\ develops, this phase can only be double-\cue.  

We also measured $M(H)$ loops for $\mathbf{H}\parallel[001]$, as shown in Figs.~\ref{MH}(d-f). This field orientation shows a similar phase transition at very similar fields, but it is sharper and more pronounced. No detwinning is observed, but none would be expected since in this case the field is at equal angles to the planes in which the spins lie in the two domains. The surprising apparent anisotropy of this transition resembles that found previously for the spin-flop transition in \two\,\cite{Adler2022}, and identifying this transition may shed additional light on the single-layer material. That detwinning occurs near this transition suggests that the higher-field phases may not twin, that strong fluctuations of the order are found near this transition, or that very different magnetic structures are obtained at higher fields.

\begin{figure}[t]
  \includegraphics[width=\columnwidth]{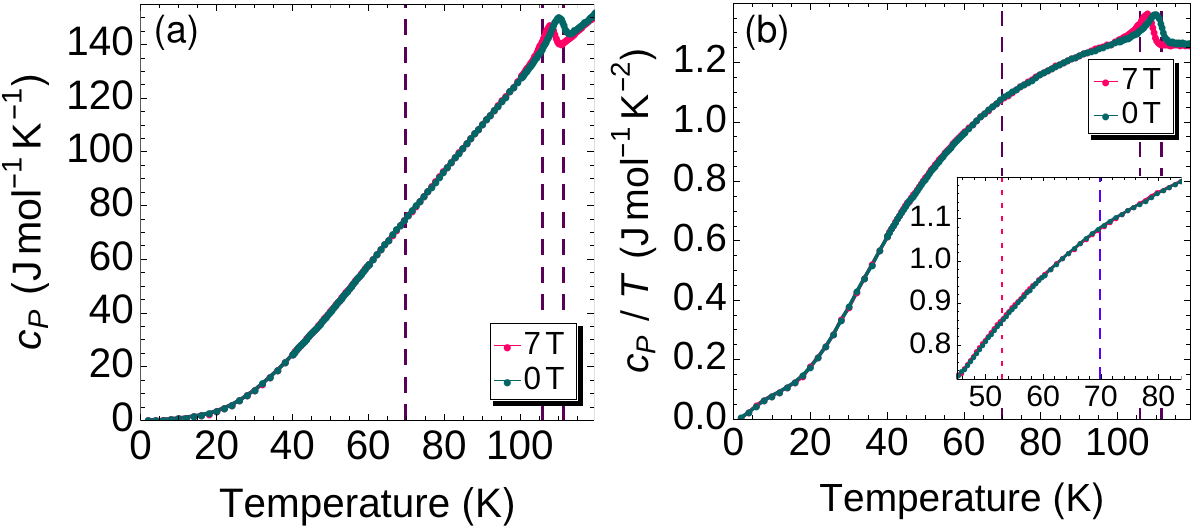}
  \caption{\label{cP}Specific-heat data on \sfo. (a) Specific heat and (b) specific heat divided by temperature for \sfo, measured in zero applied field and in a field of 7\,T along [001]. The transitions found at low field in the magnetization are marked.  The inset offers an expanded view around 70\,K, where the transition suppressed by 7\,T is also marked.}
\end{figure}

\begin{figure*}
  \includegraphics[width=\textwidth]{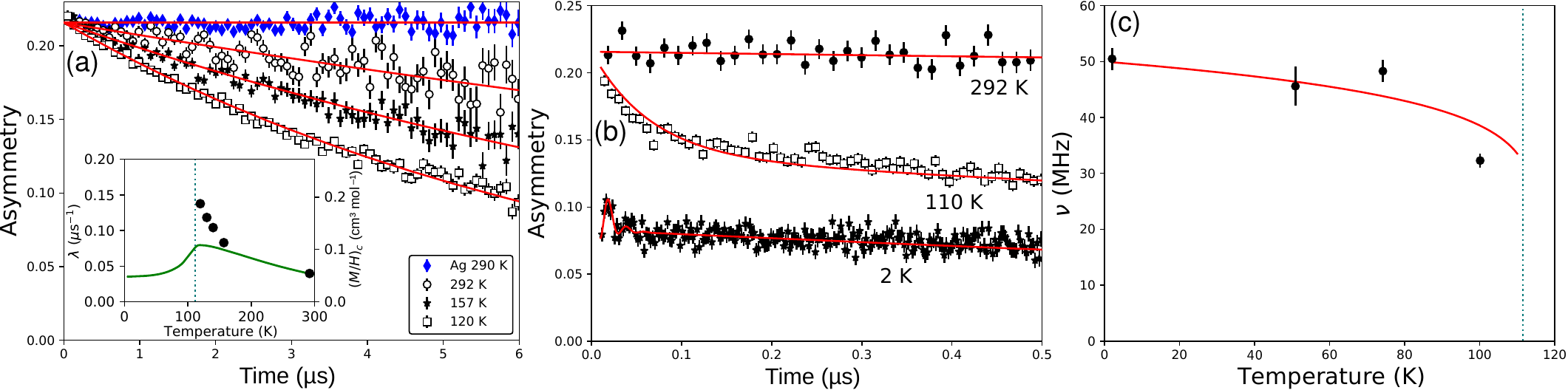}
  \caption{\label{muSR}$\mu$SR results on \sfo. $\mu$SR asymmetry in zero applied field in (a) the paramagnetic state, and (b) the magnetically ordered state. Data on high-purity nonmagnetic Ag is included as a baseline.  (c) Temperature dependence of the oscillation frequency, compared against scaled M\"o{\ss}bauer data\,\cite{Kuzushita2000}.}
\end{figure*}

\section{Specific Heat}

Since clear transitions are seen in the magnetization and dilatometry data below \TN, the specific heat was measured to determine the entropy associated with these transitions. As can be seen in Fig.~\ref{cP}, there is no clear signature of additional thermodynamic phase transitions below \TN. This indicates that the additional transitions are either broad crossovers or are associated with very small changes in entropy.  In particular, there is clearly no spin species or spin component that orders or disorders at these transitions.  The $c_P/T$ suggests a buildup of entropy below 20\,K, presumably magnetic, perhaps associated with the freezing transition seen in the difference between ZFC and FC magnetization. This did not respond to a field of 7\,T along [001].

\section{Muon Spin Rotation}

The implanted muon is a very sensitive local magnetic probe, which in particular can demonstrate clearly whether the helical order becomes commensurate in any of the magnetic phases. In zero applied field in a magnetically ordered solid, the muon experiences a spontaneous field from the magnetic order. The muon spin precesses about any transverse component of this field, and, in the simplest case, the Fourier spectrum has a single resonance at the corresponding Larmor frequency. In a helimagnet with a long pitch or incommensurate wavevector, muons stopping at different positions along the helix experience different local fields, and in the continuous limit, the spectrum approaches a broad sinusoidal distribution\cite{Guguchia2021}. The local field distribution is not very sensitive to the precise ordering wavevector or details of the ordered structure. It is, however, a volume average probe that can reveal phase separation phenomena\cite{Frandsen2016} that may be difficult to detect by other means.

In our data in zero applied field, the muon spin relaxes slowly in the paramagnetic state --- see Fig.~\ref{muSR}(a) --- however, the relaxation appears exponential rather than the Gaussian expected from nuclear dipoles, which are static on the timescale of $\mu$SR. In fact, there are few nuclear moments in \sfo, the most important being the $\sim 7$\%-abundant $^{87}$Sr. %, consequently nuclear spin dipolar relaxation will be very slow. 
The exponential relaxation should thus be due primarily to the fluctuating fields of the Fe moments. This is confirmed by the temperature dependence of the relaxation rate $\lambda$ obtained from single-exponential fits [Fig.~\ref{muSR}(a) inset], which shows a clear increase as the Fe spins slow on the approach to the N\'eel transition. The temperature dependence $\lambda(T)$ is stronger than the bulk static uniform magnetic susceptibility (green curve: $\mathbf{H}\parallel [001]$, 0.2\,T). This is unsurprising, since $\lambda$ is a local property and determined by an integral over all $q$, including the ordering wavevector, while the $q=0$ response will be suppressed by the occurrence of strong antiferromagnetic correlations in the paramagnetic state.

Below \TN\ the magnetic order gives rise to a static internal field at the muon site, changing the relaxation dramatically as seen in Fig.~\ref{muSR}(b).  Deep in the ordered state at 2\,K, a large internal field causes rapid precession of a large fraction of the spin polarization. However, this precession is nearly invisible due to extremely rapid relaxation of the spontaneous oscillations. At such a low temperature, the relaxation is probably also {\itshape static} in nature, reflecting a broad distribution of internal fields.  This is consistent with helimagnetic order for all temperatures below \TN. Measurements in an applied transverse field (not shown) confirm that the full volume is magnetically ordered. 

Fitting the rapidly damped oscillations, which are confined to the first $\sim$100\,ns of data, reveals a frequency (field) of roughly 50\,MHz (0.37\,T) at low temperature. Although this field is quite large, it is much smaller than the fields seen by M\"ossbauer spectroscopy at the $^{57}$Fe nucleus\,\cite{Kuzushita2000}, but this is expected due to the much stronger hyperfine coupling in the latter. The temperature dependence of the fitted frequency is shown in Fig.~\ref{muSR}(c) together with a curve scaled from the M\"ossbauer data. The muon frequencies are roughly consistent with the M\"ossbauer temperature dependence, confirming that the internal field (proportional to the ordered moment) rises rapidly below \TN. The rapid damping of the precession reflected in the large error bars and scatter in Fig.~\ref{muSR}(c) precludes detection of more subtle distinguishing features of the ordered phases.

\section{Phase Diagrams}

\begin{figure}[t]
  \includegraphics[width=\columnwidth]{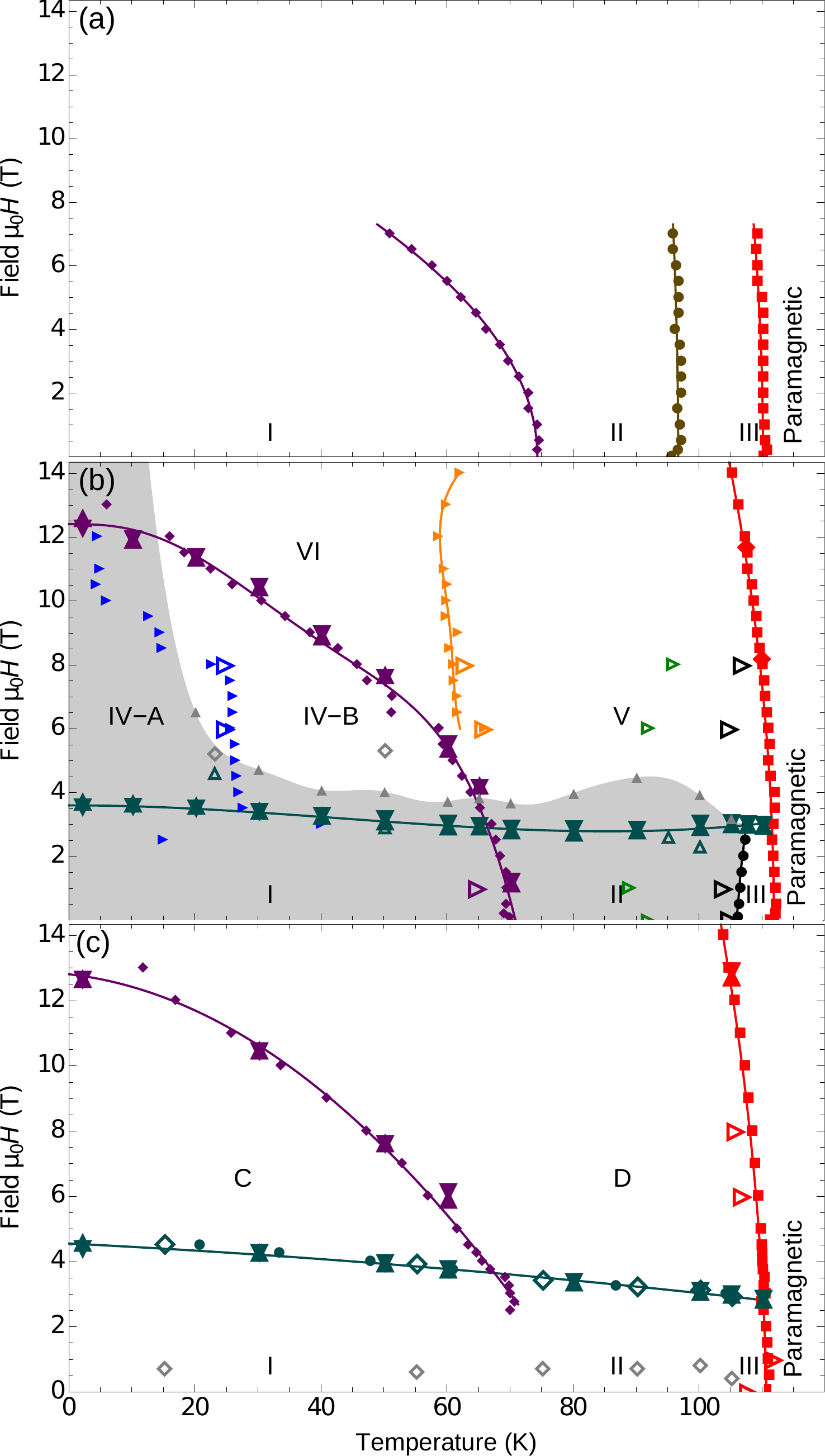}
  \caption{\label{PhaseDiag}Phase Diagrams. $H$--$T$ phase diagrams for \sfo\ extracted from the magnetization (closed symbols) and dilatometry data (open symbols) for fields along (a) [100], (b) [110], and (c) [001].  Triangles point in the direction of the field or temperature sweep, other symbols represent transitions that do not depend on sweep direction. Field sweeps were not performed for $\mathbf{H}\parallel[100]$, so these data were not sensitive to the 3--4\,T transition. Shading in (b) indicates the approximate region in which field training can be discerned.}
\end{figure}

It is possible to extract phase transitions from the magnetization and dilatometry data, most readily from extrema in their derivatives, to generate $H$--$T$ phase diagrams for various field orientations.  Our data allow us to present such phase diagrams for $\mathbf{H}\parallel [100]$ in Fig.~\ref{PhaseDiag}(a), $\mathbf{H}\parallel [110]$ in Fig.~\ref{PhaseDiag}(b), and $\mathbf{H}\parallel [001]$ in Fig.~\ref{PhaseDiag}(c). Several features only appear under field-training, which was only possible for [110] fields, or under zero-field-cooling conditions, which were not measured for $\mathbf{H}\parallel [100]$, and field sweeps were also not measured for [100], so the [100] and [001] phase diagrams should be viewed as incomplete. However, there are some surprising similarities. In particular, the transition at $\sim$3\,T is nearly isotropic, and the transition that starts at 70\,K is suppressed by field in a nearly identical manner, independent of field orientation. Isotropic phase transitions are not expected in a highly anisotropic layered crystal lattice, or in light of the previously reported elliptical helix propagating along (110)~\cite{Kim2014}. The decrease in magnetization at the 70\,K transition is comparable to that at \TN, and the change in slope in $M(H)$ around 3\,T is a factor of 2 at many temperatures and is clearly seen in dilatometry, indicating that these are unambiguously intrinsic, bulk transitions. That the former is not clearly seen in the specific heat indicates that it is either a broad crossover or not associated with a large change in entropy.  Perhaps these transitions correspond to energy scales in the magnetic interactions or spin reorientations, but detailed diffraction studies in field, and for different field orientations, are required to clarify this issue. A weak suppression of \TN\ for any field direction is less surprising, since an  applied field will eventually destabilize helical order.

Dilatometry points are shown in the phase diagrams as open symbols, with triangles pointing in the direction of the field or temperature sweep and diamonds used for magnetostriction transitions which were consistent for both sweep directions. These points largely agree with those from magnetization, as already discussed, but there are a few inconsistencies.  In particular, the boundary between phases I and II evolves into the VI-V boundary in the dilatometry measurements, rather than the IV-V boundary. There are also points around 90\,K for $\mathbf{H}\parallel [110]$ and around 0.75\,T for $\mathbf{H}\parallel [001]$ which do not correspond to features in the magnetization. The latter is evidently not related to the magnetic order since it also appears at 0.8\,T at 140\,K in the paramagnetic phase, while the former could conceivably be structural in origin.

Shading in Fig.~\ref{PhaseDiag}(b) indicates the approximate maximum extent of field-training, based on a judgement of to what field the last vestiges of this effect can still be observed in $M(H)$ and its derivative (gray triangles). This onset of field training corresponds roughly to the onset of a difference between FC and ZFC data at low temperature, and to the 3--4\,T transition at intermediate temperatures. No field training is observed in phase III, and it is unclear whether phase V supports twinning, but the inability of a 14\,T field to detwin the magnetism up to 10\,K implies that phase IV and presumably phase VI can twin.  These phases can be detwinned at higher temperatures or by cooling into them in field.

With only two directions for the propagation vector, it is difficult to produce three distinct combinations to explain the phases at zero field.  Possibilities include a subtle structural change due to magnetoelastic coupling, order of orbitals or charge multipoles, temperature-dependent changes to the propagation vector, ordering of an overlooked component of the spin, or some form of exotic multiple-\cue\ order such as those proposed theoretically in other contexts in Refs.~\onlinecite{Wang_2021,Okumura2020,Hayami2022,Yambe_2301.12629} but not yet demonstrated experimentally.

A subtle orthorhombic distortion associated with charge order in \sfo\ is observed below $\sim$330\,K, but with an extremely short correlation length along the $c$ direction~\cite{Kim2021}. The sharp magnetic Bragg reflections in both single-crystal and powder diffraction imply a long magnetic correlation length in all directions. This means that every magnetic domain will average over many structural domains, and the material will be effectively tetragonal from the point of view of the magnetism. This is particularly true once the magnetism is detwinned and the entire sample is a single magnetic domain.  The iron is close to Fe$^{3+}$ (high spin), which has no orbital polarization and is spherically symmetric.  Its excess positive charge is predominantly delocalized on the oxygen cages, which should preclude any orbital or charge multipole ordering. Any significant magnetoelastic coupling should make the transitions within the ordered state visible in the specific heat data, particularly in field, which they are not, and a structural component to the transition would have been seen in Ref.~\onlinecite{Kim2021}.  These transitions are presumably magnetic.  We thus investigated the temperature and field dependence of the propagation vector.

\section{Magnetic Order}

Looking first at the low-field phases, we note that the magnetic state of \sfo\ has been previously reported as an elliptical helix based on neutron diffraction~\cite{Kim2014}.  Since we have identified an unexpectedly complex phase diagram for all field orientations, one immediate question is whether there is any obvious change in the propagation vectors, ellipticity, or intensities of magnetic reflections, which would give a hint as to the nature of the magnetic phase transitions.  We have seen above that the $\mu$SR results are consistent with helical magnetism at all temperatures below \TN, so we now turn to diffraction.

\begin{figure}[t]
  \includegraphics[width=\columnwidth]{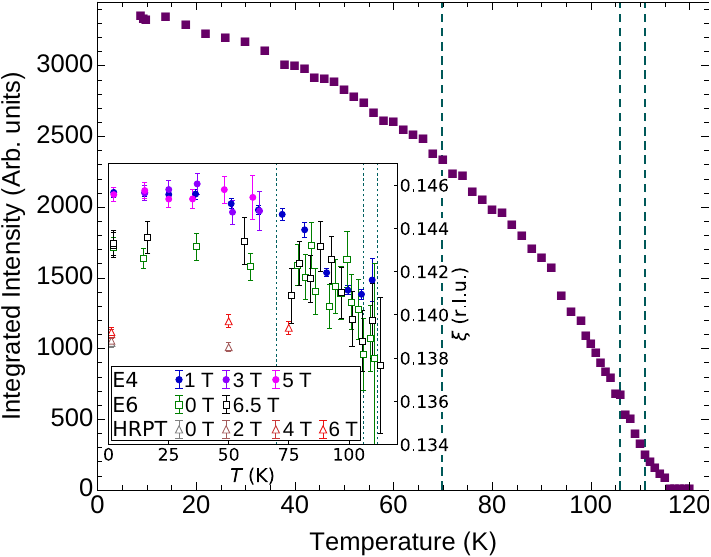}
  \caption{\label{E5}Temperature- and field-dependent incommensurability from neutron diffraction. Diffracted magnetic intensity in zero field at E5, with the transitions from the low-field $\mathbf{H}\parallel[110]$ magnetization marked.  The integrated intensity in the ($\xi$\,$\overline{\xi}$\,$\overline{1}$) reflection shows no signature of a transition within the magnetically ordered state.  Inset: the incommensurability is reduced slightly on warming toward \TN, and appears insensitive to the magnetization transitions.}
\end{figure}

\begin{figure}[tb]
    \includegraphics[width=\columnwidth]{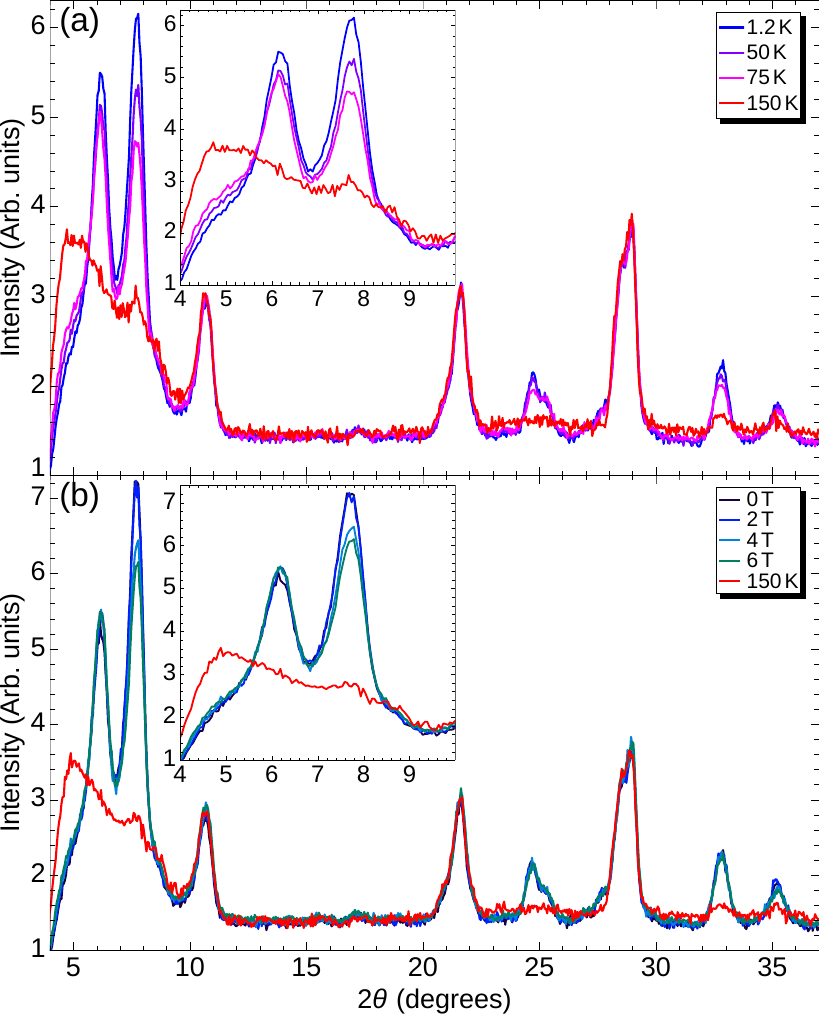}
    \caption{\label{neutron_peaks}Effect of temperature and field on magnetic Bragg reflections. Evolution of magnetic neutron intensity (HRPT) with (a) temperature and (b) magnetic field. Insets highlight the strongest magnetic peaks for an impurity phase of \cubic\ (6.1$^\circ$) and \sfo\ (7.7$^\circ$). Datasets in the paramagnetic phase at 150\,K and 0\,T (a) or 6\,T (b) are included for reference.}
\end{figure}

The diffracted intensity in zero applied magnetic field was tracked versus temperature for a single-crystal sample at E5 and for powder samples at E6 and HRPT [the latter is shown in Fig.~\ref{neutron_peaks}(a)], and the magnetic Bragg peaks remain at their incommensurate positions.  As shown in Fig.~\ref{E5}, there are no sharp changes in the intensity of the magnetic reflections with temperature, and in particular there is no signature of the transitions found in the magnetization. The temperature and field dependence of the incommensurability measured on powder samples at E6 and HRPT and a single crystal measured at E4, shown in the inset to Fig.~\ref{E5}, are smooth and on the scale of variations among samples or beamlines.  That the incommensurability appears to be insensitive to all magnetization transitions indicates that there are no significant changes to the underlying \cue\ vectors with temperature or with magnetic field up to at least 6.5\,T.  This suggests that, as in \cubic, these phases are distinguished by different combinations of \cue\ vector; however, as mentioned above, \sfo\ has three magnetic phases at low field and at least two more above $\sim$3\,T, but a maximum of only two independent \cue\ vectors.  It remains unclear what distinguishes phases I and II.  

To investigate how the higher-field phases differ from the low-field phases, diffraction was performed in magnetic fields $\mathbf{H}\parallel[001]$ and [110] at E4 and on powder at HRPT\,---\,the latter is shown in Fig.~\ref{neutron_peaks}(b).  The volume of reciprocal space blocked by the magnet (E4 and E6), the random field orientation on the powder sample (HRPT), and possible field-induced preferred orientation effects all limit what can be said about the high-field phases, but the changes in position of the magnetic reflections were again minimal, as seen in Fig.~\ref{neutron_peaks}(b) and summarized in the inset to Fig.~\ref{E5}.  Intensities in the magnetic peaks changed across the 3--4\,T transition in a manner suggestive of a reduction in the in-plane component of the ordered moment.  Based on the previously reported elliptical helix~\cite{Kim2014}, this would indicate a higher ellipticity.  However, the change across this transition appears to be relatively abrupt and step-like, and it remains unclear why the ellipticity should be quantized.  Clarifying the nature of the higher-field phases and their relationship to the low-field phases, as well as fully identifying the low-field phases, will require a detailed single-crystal diffraction study in a magnet capable of applying at least 4\,T along [110], to detwin the low-field phases and access the high-field phases.  We note, however, that a similar-looking transition at 3--4\,T was found in magnetization data on powders of the single-layer analogue \two, where it was reported as most likely a spin-flop transition~\cite{Adler2022}.  This transition is presumably also relatively isotropic, despite the strong structural anisotropy. \two\ is only available in powder form, so clarifying the nature of the 3--4\,T transition in \sfo\ will likely also provide strong hints as to the magnetic phase diagram of \two.

Our inability to detwin phase III makes it a prime candidate for double-\cue\ order analogous to the skyrmion-lattice phase I$_\text{c}$\,\footnote{Since the magnetic phases in \cubic\ use the same lettering as used here, we use a subscript ``c'' to distinguish the phases in cubic \cubic.} in \cubic.  In contrast, both phases I and II in \sfo\ could be detwinned, indicating that they both break the four-fold rotational symmetry of the lattice.  Any multiple-\cue\ order in either of these phases would need to be extremely exotic, but it also remains unclear how to realize two independent single-\cue\ phases with helical order alone.  The ordering of an overlooked spin component would be possible, particularly in phase I, since the loss of this order would be expected to enhance $M/H$ on warming.  However, previous refinements of the magnetic order were performed on single crystals at low temperature, and should have detected this.  The helical order at low temperature has been reported to be elliptical~\cite{Kim2021}, so the ellipticity could change, as suggested across the 3--4\,T transition, but no clear change is seen with temperature.  

Above the 3--4\,T transition, while it is possible to freeze the magnetic order at low temperatures and prevent detwinning (distinguishing phase IV-A from IV-B), the higher-field phases otherwise seem to be largely detwinned.  The peak in the magnetization separating phases V and VI could perhaps arise from fluctuations as the magnetic order reorients itself in some way.  However, we have not observed a clear change across the IV-V boundary with diffraction, and our magnetic fields were not high enough to access phase VI, so differences among the higher-field phases remain unclear.  Identifying these phases will require detailed high-field diffraction measurements on single crystals.

\section{Summary and Outlook}

The magnetic phase diagram of \sfo\ is surprisingly complex, and highly reminiscent of that of \cubic.  This is despite \cubic\ having four distinct directions for its propagation vector pointing along \{111\}, while there are only two such directions possible in \sfo.  The high-temperature phase III cannot be detwinned by field, making it evidently a double-\cue\ phase, possibly analogous to the low-temperature skyrmion-lattice phase I$_\text{c}$ in \cubic.  However, it remains unclear what distinguishes phases I and II.  The transition at 3--4\,T, likely analogous to the ``spin-flop'' transition in \two~\cite{Adler2022}, may be related to the ellipticity of the helical order.  The other transitions and the identities of the remaining phases remain unclear.  The phase diagram of \cubic, despite some similarities, provides limited insight here\,---\,its quadruple-\cue\ phase II$_\text{c}$ is impossible in \sfo, and its phases III$_\text{c}$, IV$_\text{c}$, and V$_\text{c}$ have not been identified.  At higher fields, there is very little diffraction data on either material to provide insight.  Since \sfo\ has only two possible propagation directions for its helical order, with spin orientations in orthogonal planes, perfect detwinning of the magnetic order is possible, and we have shown that this is readily achieved at accessible temperatures and fields.  This is in contrast to \cubic, in which it is not possible to fully detwin all magnetic phases with a magnetic field.  Fully determining the magnetic phases in \sfo\ will be more straightforward and is likely to provide insight for \cubic, allowing better targeting of future measurements as that material's phase diagram is elucidated.  

The single-layer analogue \two\ is possibly more relevant to the current work, but less is known of its magnetic structures.  This is largely because it decomposes far below the liquidus~\cite{Fossdal2004}, making crystal growth impossible thus far.  A spin-flop transition reported in that material in field~\cite{Adler2022}, which must be relatively isotropic since this was measured on powder, closely resembles the 3--4\,T transition seen here.  In \sfo\ this transition appears to be connected with a relatively sharp change in the ellipticity of the helical order, but such a relatively abrupt change in a parameter which ought to be continuous is surprising, suggesting that our understanding of the low-field phases is incomplete.  Diffraction on single crystals should be performed to nail down the phases in \sfo, which will in turn allow inferences as to the magnetic phase diagram of \two.

It is worth commenting here that while \cubic\ is too symmetric to support DM interactions, the charge disproportionation in \sfo\ should lead to a lattice distortion which would allow them. Yet, the strong similarities in the magnetic order and phase diagrams among the three better-studied members of this family indicate that DM interactions play no significant role. We would thus anticipate a similar phase diagram and similar magnetic order in the triple-layer analogue \four, which to our knowledge has not been investigated. The helical and multiple-\cue\ order found in \sfo\ and \two, and likely also present in \four, must arise from the same competition among exchange interactions, without DM, even if DM interactions are allowed.

%  Anita Fossdal's thesis should include 4-3-10.
%  https://bibsys-almaprimo.hosted.exlibrisgroup.com/permalink/f/1bgnrh8/BIBSYS_ILS71481177140002201
%  It is not readily accessible without being in Norway.
%  However, this is an engineering thesis, and there is no evidence that anything was cooled below room temperature.  

%  Sandra Dann's thesis does not report the 4-3-10.

In light of its surprisingly complex magnetic phase diagram, \sfo\ calls for more detailed investigation to identify its magnetic phases and phase transitions.  The diffraction, in particular, should be revisited at high fields and under field-trained conditions, and transport properties may reveal signatures of topological protection that would help clarify which phases are multiple-\cue.  It would also be worth revisiting the [100] and [001] phase diagrams in a vector magnet, which would allow field-training into a single-domain state before measuring.  While it is not yet possible to identify most of the magnetic phases found in \sfo, its magnetic phase diagram is clearly much richer than previously imagined, and it will likely yield several exotic magnetically ordered phases.  

\begin{acknowledgments}
The authors are grateful for experimental assistance from the groups of M.\ Jansen and R.\ Kremer.  This project was funded by the Deutsche Forschungsgemeinschaft (DFG, German Research Foundation) through individual grants PE~3318/3-1 (Project No.\ 455319354), IN~209/7-1 (Project No.\ 401179363), and IN~209/9-1 (Project No.~434257385); through projects C01 and C03 of the Collaborative Research Center SFB~1143 (Project No.\ 247310070); through the W\"urzburg-Dresden Cluster of Excellence on Complexity and Topology in Quantum Materials\,---\,\textit{ct.qmat} (EXC~2147, Project No.\ 390858490); and through the Collaborative Research Center TRR~80 (Project No.\ 107745057). Neutron measurements were carried out at the E4, E5, and E6 instruments at the BER-II research reactor, operated by the Helmholtz-Zentrum Berlin f\"ur Materialien und Energie. This work is based in part on experiments performed at the Swiss spallation neutron source SINQ, Paul Scherrer Institute, Villigen, Switzerland. The authors also acknowledge the Institut Laue-Langevin, Grenoble (France) for providing neutron beam time\,\cite{ILL_4-01-1589_IN5}.
\end{acknowledgments}

\bibliography{327_MH}

\end{document}